\documentclass[prb,aps,reprint,twocolumn]{revtex4}

\usepackage{graphicx}
\usepackage{amsmath}
\usepackage{amsfonts}
\usepackage{amssymb}
\usepackage{upgreek}

\usepackage{epstopdf}
\usepackage[usenames,dvipsnames]{xcolor}

\renewcommand{\imath}{i}

\DeclareMathOperator{\ch}{ch}
\DeclareMathOperator{\sh}{sh}

\begin{document}

\title{Diffusion-induced dissipation and mode coupling in nanomechanical resonators}
\author{Christin Edblom}
\author{Andreas Isacsson}
\email{andreas.isacsson@chalmers.se}

\affiliation{Department of Applied Physics, Chalmers University of Technology, 
					S-412 96 G\"oteborg, Sweden}

\date{\today}

\begin{abstract} 
	We study a system consisting of a particle adsorbed on a carbon nanotube 
	resonator. The particle is allowed to diffuse along the resonator, in order 
	to enable study of e.g. room temperature mass sensing devices. The system is 
	initialized in a state where only the fundamental vibration mode is excited, 
	and the ring-down of the system is studied by numerically and analytically 
	solving the stochastic equations of motion.  We find two mechanisms of 
	dissipation, induced by the diffusing adsorbate. First, short-time 
	correlations between particle and resonator motions means that the net effect 
	of the former on the latter does not average out, but instead causes non-exponential
	dissipation of vibrational energy. For vibrational amplitudes that are much 
	larger than the thermal energy this dissipation is linear; for small 
	amplitudes the decay takes the same form as that of a nonlinearly damped 
	oscillator. Second, the particle diffusion mediates a coupling between 
	vibration modes that opens a new dissipation channel by enabling energy transfer from the fundamental mode to the
	excited modes, which rapidly reach thermal equilibrium. 
\end{abstract}

\maketitle

%%%%%%%%%%%%%%%%%%%%%%%%%%%%%%%%%%%%%%%%%%%%%%%%%%%%%%%%%%%%%%%%%%%%%	

\section{Introduction}
Nanoelectromechanical (NEM) resonators hold great promise for
applications in inertial mass sensing~\cite{Roukes_2011,
  Roukes_2012}. Carbon nanotubes (CNTs) in particular are suited when
striving for high sensitivity~\cite{Bachtold_2008,Zettl_2008}, due to
their extremely low mass. Recently, using CNT resonators, yoctogram
sensitivity was achieved in experiments~\cite{Bachtold_2012}. In mass
sensing applications, it is commonly assumed that an adsorbate, once
attached to the surface, remains in the same positions during the time
of measurement. However, at elevated temperatures thermal fluctuations
can cause the adsorbate to change its position along the tube via
diffusion.  As the resonant frequency of the system depends on the
position of the adsorbate, this gives rise to frequency fluctuations
with accompanying phase noise.

For driven resonators, the effect of such frequency fluctuations has
recently been studied both theoretically~\cite{Dykman_2010,
  Atalaya_2011, Atalaya_2011_2, Atalaya_2012} as well as
experimentally~\cite{Roukes_Diffusion}. The effect manifests in a
broadening and/or changed shape of the resonant response. However,
broadening also arises from dissipation of mechanical energy.
Dissipation and the origin of Q-factor limitations in
nanoelectromechanical systems has been a long standing research topic
where there are still unresolved issues~\cite{Lifshitz_2000,
  Lifshitz_2001, WilsonRae_2008, Eom_2011}. Recently, the connection
between dissipation and nonlinear phenomena in NEM-resonators has begun to attract
attention. This is partly because of the presence of nonlinear
damping~\cite{Eichler_2011, Croy_2012} in carbon nanoresonators, and
partly due to the recognition that geometric nonlinearites themselves
give rise to dissipation~\cite{Midtvedt_2014} and spectral
broadening~\cite{McEuen_PNAS}.  While it was shown in
Refs.~\onlinecite{Atalaya_2011} and~\onlinecite{Atalaya_2011_2} that a
diffusing particle on an otherwise linear resonator induces both
spectral broadening as well as a nonlinear response to driving, we also expect the same mechanism to give rise to dissipation and mode coupling. In that case, two questions arise; in what manner does the system relax to equilibrium, and what is the effect of the mode coupling on the the system dynamics? In the present study, we investigate the characteristics of the dissipation process induced by the diffusing adsorbate in order to answer these questions. 

\begin{figure}[t]
\centering
\includegraphics[width=\linewidth]{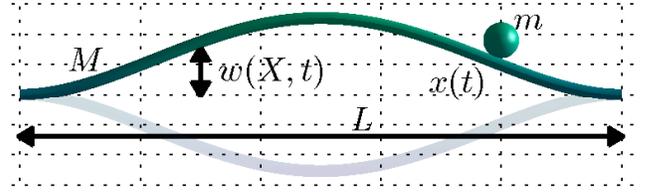}
\caption{(color online) A particle of mass $m$ is adsorbed on a
  one-dimensional resonator of mass $M$ and length $L$. The transverse
  displacement of the resonator at coordinate $X$ along the nanotube
  axis is $w(X,t)$; $x(t)$ is the position of the adsorbed
  particle. The resonator is initialized in its fundamental vibration
  mode, and the effect of the stochastically diffusing particle on the
  ring-down of the resonator is studied.\label{fig:system}}
\end{figure}

For concreteness, we model a diffusing adsorbate along a carbon
nanotube-resonator and study the decay of mechanical
energy in the system. 
In order to isolate the dissipative contribution to the resonance
broadening, we simulate ring-down measurements such as the ones
performed in Refs.~\onlinecite{Meerwaldt_2013}
and~\onlinecite{Leeuwen_2014}.  Our model is a doubly clamped
one-dimensional resonator constrained to move in the plane. A small
mass is adsorbed on the resonator, and allowed to diffuse along it as
shown in Figure~\ref{fig:system}.  The resonator is excited in its
lowest flexural vibration mode, by means of e.g. a nearby gate, and
the subsequent free evolution of the system is studied.

\begin{figure}[t]
\centering
\includegraphics[width=\linewidth]{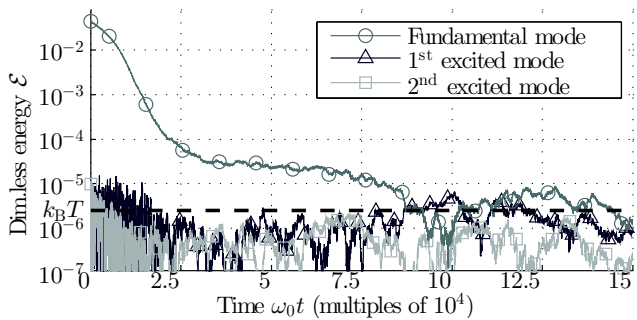}
\caption{(color online) Dimensionless mode energies as function of time during 
  ring-down of the fundamental mode. Here $\epsilon\simeq10^{-2}$ and $T=500$~K. For 
  clarity, only the lowest lying flexural modes are shown;
  higher modes behave similarily. The dashed line indicates the
  thermal energy in dimensionless units.  The system is initialized in
  a state where all energy ($\mathcal E_0(0)\simeq 10^4 k_{\mathrm
    B}T$) is in the fundamental mode, and then allowed to evolve
  freely. As can be seen, the effect of the particle diffusion is to
  damp out the fundamental mode and establish equilibrium with
  higher-lying modes. 
\label{fig:decay}}
\end{figure}

An example of the distribution of mechanical energy between flexural modes during a simulated ring-down experiment is shown in Figure~\ref{fig:decay}. A non-exponential decay of the mechanical energy of the fundamental mode is evident; eventually 
thermal equilibrium is reached. In Section~\ref{sec:singlemode} we show that, in the limit of a single flexural mode, the observed decay can be divided into two distinct regimes. In the first regime the vibration amplitude is large, the adsorbate is trapped at an antinode of the vibration, and the mechanical energy decays linearly. In the second regime the amplitude is small and the adsorbate diffuses freely along the nanotube, which exhibits non-exponential dissipation characteristic of nonlinearly damped resonators. This damping is due to the fact that the inertial force acting on the particle causes its
motion to have frequency components twice that of the fundamental
mode. Because of retardation, this short-time correlation between
adsorbate motion and resonator motion causes dissipation: an effect also seen in
molecular dynamics studies on graphene resonators~\cite{Park_2014}. 
In addition, as the particle changes position, a
mode coupling is induced that opens a new channel of dissipation, allowing the transfer of energy to higher-lying modes and
causing them as well to equilibrize. As we discuss in Section~\ref{sec:multimode}, this new dissipation channel means that while the results of Section~\ref{sec:singlemode} are qualitatively robust, a quantitative error arises; the rate of decay due to the mode coupling is linearly proportional to the number of higher modes (up to a parameter-dependent cut-off), and hence their total energy. This result is analogous to those found in Refs.~\onlinecite{Midtvedt_2014} and~\onlinecite{McEuen_PNAS} when introducing conservative geometric nonlinearities in clean nanoresonators.

%%%%%%%%%%

\section{Equations of motion for a resonator with a diffusing particle}
As shown in Figure~\ref{fig:system}, we consider a resonator of length
$L$ with mass $M=L\rho$ and bending rigidity $\kappa$. Neglecting
longitudinal displacement, the Lagrangian density for the unperturbed
resonator is~\cite{LL}
\begin{eqnarray}
&&\mathcal L_0=\frac12\rho\dot w^2-\frac12\sigma w_X^2-\frac 1 2 \kappa w_{XX}^2.
\label{eq:unpertL}
\end{eqnarray} 
Here, $w=w(X,t)$ is the transverse displacement (see
Figure~\ref{fig:system}), $X$ is the coordinate measured along the
resonator, and $w_X=\partial w/\partial X$. In the limit of small vibration amplitude and/or
large prestrain, the built-in tension $\sigma$ can be assumed
independent of $w$. The unpertubed Eigenfrequencies $\omega_n$ and
Eigenmodes $\phi_n(X)$ are found from solving the corresponding
equation of motion, see appendix~\ref{sec:eigenmodes}.  For convenience, we will
 work with eigenmodes normalized so that $\int dx \phi_n\phi_m=L\delta_{nm}$, 
and with boundary conditions corresponding to a doubly clamped beam: 
$w(0)=w(L)=w_X(0)=w_X(L)=0$.

Including the adsorbate of mass $m=\epsilon M$ at $X=x(t)$, the total
Lagrangian is
\begin{eqnarray}
&&\mathcal L=\mathcal L_0+\frac12m\delta(x-X)\left(\dot x^2+\left(\dot
  w+\dot xw_X\right)^2\right).
\end{eqnarray}
Here, the term added to the resonator Lagrangian density $\mathcal L_0$ is the kinetic energy $\frac12 m\dot{\mathbf r}^2$ of the adsorbed particle, where its position $\mathbf r(t) = \big(x(t), w(x(t),t)\big)$. 

We expand the displacement in Eigenmodes, $w(X,t)=\sum_n
q_n(t)\phi_n(X)$, and by variation of $\mathcal L$ we find the equations of motions 
\begin{eqnarray}
&&\ddot {q}_n+\omega_n^2q_n-\epsilon \phi_n(x)\sum_k\omega_k^2q_k\phi_k(x)=0, \label{eq:eom1}\\
&&\dot{x}-\frac1\gamma\sum_{k,\ell}\omega^2_kq_kq_{\ell}\phi_k\phi_\ell'=\sqrt{D}\eta(t).\label{eq:eom2}
\end{eqnarray}
In order to allow for the thermal diffusion of the adsorbate, a stochastic force has been introduced in the right-hand side of Equation~\eqref{eq:eom2}. By the fluctuation-dissipation theorem, this force is accompanied by a damping rate~$\gamma$. Thus, $\eta(t)$ is a delta-correlated Gaussian noise,
i.e. $\left<\eta(t)\eta(t')\right>=\delta(t-t')$, and $D={2k_{\mathrm
    B}T} /{m\gamma}$. Throughout, we assume reflecting boundary
conditions for the diffusing particle. Finally we note that Equations~\eqref{eq:eom1}-\eqref{eq:eom2} are derived using the approximation $\partial_t^2w(x(t),t)\approx-\sum_n\omega_n^2q_n(t)\phi_n(x)$, which is equivalent to claiming that the effect of the added mass is a small correction to the unperturbed motion, as well as an assumption of strong damping ($m\ddot x\ll m\gamma\dot x$) that allows the inertial term to be neglected.

\begin{figure}[t]
\centering
\includegraphics[width=\linewidth]{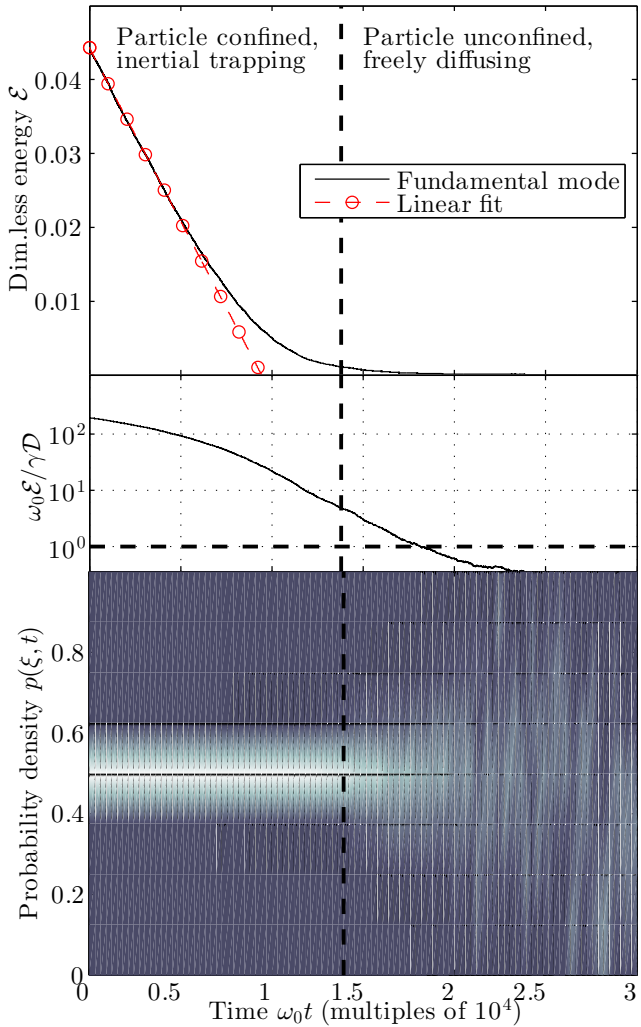}
\caption{(color online) Top: fundamental mode energy as a function of
  time, together with a linear fit to the initial decay. Center: the
  parameter $\omega_0\mathcal E/\gamma\mathcal D$, shown below to
  govern the qualitative behavior of the system, as a function of
  time. The horizontal dashed line indicates where $\omega_0 \mathcal
  E=\gamma\mathcal D$.  Bottom: evolution of the probablity
  distribution $p(\xi,t)$, where $\xi=x/L$, for the particle to be at
  a certain position $0<\xi<1$ along the resonator as a function of
  time. For large initial amplitudes, the particle remains inertially
  trapped at the antinode of the vibration around $\xi \approx
  0.5$. The corresponding energy decay is linear in time. As the
  thermal fluctuations overcome the inertial trapping potential, the
  particle diffuses freely, and the energy decays algebraically towards
  equilibrium. \label{fig:confine}}
\end{figure}

The non-linear system of Equations~\eqref{eq:eom1}-\eqref{eq:eom2} is numerically 
integrated using a second-order algorithm~\cite{Manella_1989}. The system is initialized in a state
where all energy is stored in the fundamental mode, $q_n(t=0)=0,
n>0$, and the particle is adsorbed at $x=L/2$. The resonator 
dimensions have been chosen to be experimentally realistic:
length $1\ \upmu$m, diameter 5~nm, and fundamental resonant frequency
$\omega_0=2\pi\times108$ MHz. 
Some results are 
shown in Figures~\ref{fig:decay} and~\ref{fig:confine}, using parameters ${\cal  D}
=D/\omega_0L^2=2.85\times10^{-4}$, $\epsilon=1.82\times 10^{-2}$ (corresponding to an adsorbate mass $m$ similar to that of a mid-sized protein molecule or a larger collection of non-interacting smaller adsorbates), and $\gamma=0.241
\omega_0$. The initial amplitude
$q_0$ is here chosen such that $\mathcal E_0(t=0)=0.044$, where
$\mathcal E_n=\omega_n^2q_n^2/ \omega_0^2L^2$ is the dimensionless
kinetic energy stored in mode $n$. 

The main features to note are the
initially linear decay of energy in the fundamental mode, the initial
trapping of the adsorbate near the center of the resonator followed by
diffusion along the length, and the eventual thermalization of all
vibration modes. Also note that the higher-lying modes reach an internal equilibrium very
rapidly compared to the slow decay of the fundamental mode energy. These results are further discussed in Section~\ref{sec:multimode}. 

As illustrated in Figure~\ref{fig:confine}, one can identify two
distinct limiting cases. The first, high-amplitude limit is
characterized by the particle being trapped at the antinode of the
vibration around $x\approx L/2$. In this regime, the energy of the
resonator decays linearly in time. As the amplitude of the resonator
vibrations decreases, thermal fluctuations overcome the inertial
trapping potential, the particle starts to diffuse along the entire
length of the nanotube, and the decay rate is no longer linear.

The regime is determined by the parameter
$\omega_0\mathcal E_0/\gamma \mathcal D=\frac
12m\omega_0^2q_0^2/k_{\mathrm B}T$: the ratio between vibrational and thermal
energy of the adsorbate. The inertial trapping potential is
proportional to the vibrational energy, so the particle remains
confined as long as $\omega_0\mathcal E_0/\gamma \mathcal D\gtrsim1$, and
diffuses freely when $\omega_0\mathcal E_0/\gamma \mathcal D\lesssim1$. This
parameter is shown as a function of time in the center panel of
Figure~\ref{fig:confine}, illustrating the agreement between
the value of $\omega_0\mathcal E_0/\gamma \mathcal D$ and the
behavior of the adsorbate. 

The diffusion constant $D$ depends on the adsorbate and resonator materials. While measurements of the diffusion constants of several elements on graphite exists it is not clear that these are applicable to diffusion along a nanotube. The values for diffusion constants and adsorbate masses used in the simulations were chosen to provide good numerical stability, facilitate comparison with results obtained from perturbation theory, and to visualize the different regimes as clearly as possible. Interestingly, as will be seen in Section~\ref{sec:singlemode}, the leading relevant parameters that determines the ringdown dynamics in the pertubative regimes are the dimensionless ratios $\epsilon\mathcal D\gamma/\omega_0$ (trapping regime, Equation~\eqref{eq:linearE}) and $\epsilon\omega_0/\gamma$ (free diffusion regime, Equation~\eqref{eq:alphaD}). Since we have $\mathcal D\gamma=2k_{\mathrm B}T/m\omega_0L^2$ by virtue of the fluctuation-dissipation theorem, the precise value of the diffusion constant is not crucial in the trapping regime, provided $D\gg(L/q_0)^2(k_{\mathrm B}T/m\omega_0)$. The ratio $\epsilon\omega_0/\gamma$ for the free diffusion regime becomes $\epsilon m\omega_0L^2 D/2k_{\mathrm B}T$. Since $D$ is expected to depend exponentially on temperature, probing a large parameter range can thus be done by varying $T$.

%%%%%%%%%%

\section{Single mode dynamics}
\label{sec:singlemode}
To understand the observed energy decay we first focus on a single
flexural mode. This simplification of the equations of motion is
motivated by simulations, which have not shown any qualitative
dependence on the number of included modes. That is, even
when only the fundamental mode of the resonator is included, the two
regimes identified in Figure~\ref{fig:confine} are
evident. Below, we will treat the two regimes separately, beginning
with the large amplitude-case.

Considering only the fundamental mode,
i.e. $w(X,t)=q_0(t)\phi_0(X)$, one finds
\begin{eqnarray}
&&\ddot {q}_0+\omega_0^2[1-\epsilon \phi_0^2(x)]q_0=0, \\
&&\dot{x}=\frac{\omega_0^2}{2\gamma} q_0^2\partial_x\phi_0^2(x)+\sqrt{D}\eta(t),
\end{eqnarray}
We measure time in units of $\omega_0^{-1}$, and change to action angle 
variables $({\cal E}(t),\theta(t))$ via the
transformations $q_0(t)=L\sqrt{{\cal E}}\cos(\omega_0 t+\theta)$ and
$\dot{q}_0(t)=-\omega_0L\sqrt{{\cal E}}\sin(\omega_0 t+\theta)$. Then, the
equations take the form
\begin{eqnarray}
&&\partial_\tau{\cal E}=-\epsilon \phi_0^2 {\cal E}\sin 2\nu,\quad (\nu=\theta+\tau),\label{eq:actang1}
\\
&&\partial_\tau{\theta}=-\epsilon \phi_0^2 \cos^2 \nu,\label{eq:actang2}
\\
&&\partial_\tau{\xi}=\frac{\omega_0}{2\gamma}{\cal E}\cos^2\nu\partial_\xi \phi_0^2 +\sqrt{{\cal D}}\eta(\tau),
\label{eq:actang3}
\end{eqnarray}
where $\xi=x/L$, $\tau=\omega_0t$ and ${\cal D}={D}/{\omega_0L^2}$.
Thus, it is quite clear from Equation~(\ref{eq:actang1}) that performing an
RWA-approximation here leads to $\partial_\tau {\cal E}=0$ and that
the effect of the particle diffusion is only to cause fluctuations in
resonant frequency. Hence, in order for $\partial_\tau {\cal E}\neq
0$, $\xi$ must contain a frequency component $\sin 2\nu$ which arises
from the first term in Equation~(\ref{eq:actang3}). It follows that the observed
decay in energy stems from short-time correlations with frequency
$2\omega_0$ between particle and resonator motions. 
    
%%%%%%%%%%

\subsection{Large amplitude vibrations, confined particle}
When ${\omega_0{\cal E}}/{\gamma{\cal D}} \gg 1$, the thermal
fluctuations cannot overcome the inertial trapping potential and the
adsorbate fluctuates around the antinode of the flexural mode. In this
regime, the phase noise is typically small and can be neglected when
estimating the decay rate. Furthermore, as the particle is at all times in
the vicinity of the antinode we can make the approximation
$\partial_\xi \phi_0^2\approx
2\phi_0(0)\phi_0^{\prime\prime}(0)\xi=-k\xi$, which renders the
diffusion equation~(\ref{eq:actang3}) linear.  Solving for the
particle motion yields
$$\xi(\tau)=\sqrt{{\cal D}}\int^\tau d\tau'\, \eta(\tau')e^{-\frac{\omega_0 k}{2\gamma}\int_{\tau'}^\tau d\tau^{\prime\prime}{\cal E}(\tau^{\prime\prime})\cos^2 \nu^{\prime\prime}}.$$
Inserting back into the equation for ${\cal E}$, omitting the term vanishing upon averaging over fast fluctuations, and assuming ${\cal E}$ to be slow, one finds
\begin{eqnarray}
&&\partial_\tau{\cal E}\approx \frac12\epsilon{\cal D} k{\cal E}\sin 2\tau \int_{-\infty}^\tau d\tau'\,e^{-\frac{k\omega_0{\cal E}(\tau)}{2\gamma}\int_{\tau'}^\tau d\tau^{\prime\prime}\cos^2 \tau^{\prime\prime}}\nonumber.
\\
\end{eqnarray}
Averaging over fast oscillations, 
\begin{eqnarray}
&&\partial_\tau{\cal E}\approx \epsilon{\cal D} k{\cal E}I\left(\frac{k\omega_0{\cal E}(\tau)}{\gamma}\right)\nonumber
\end{eqnarray}
where the integral $I$ is defined as
$$I(x)=\frac{1}{2\pi}\int_0^{2\pi} d\tau \sin 2\tau \int_{-\infty}^\tau d\tau'\,e^{-x\int_{\tau'}^\tau d\tau^{\prime\prime}\cos^2 \tau^{\prime\prime}}.$$
The integral is well approximated by the expression $I(x)\approx - (4+\sqrt{\pi}
[(x/2)\coth(x/2)-1])^{-1}$. Hence, for large amplitudes such that $k\omega_0
{\cal E}(\tau)/{\gamma}\gg 1$, the decay becomes linear, i.e.
\begin{equation}
\partial_\tau{\cal E}\approx -\epsilon\frac{2{\cal D}\gamma}{\sqrt{\pi}\omega_0}.
\label{eq:linearE}
\end{equation}
Comparisons with simulation shows that this result is correct within an order of 
magnitude, even when excited modes are included: see Figure~\ref{fig:Ndep}. However, 
as discussed in Section~\ref{sec:multimode}, adding excited modes introduces more channels of decay, and consequently increases 
the decay rate.

%%%%%%%%%%

\subsection{Small amplitude vibrations, unconfined particle\label{sec:whatever}}
If $\omega_0{\cal E}/({\gamma {\cal D}})\ll 1$, the
system~(\ref{eq:actang1})-(\ref{eq:actang3}) can be solved by means of
perturbation theory. This limit can be seen to be equivalent to the
assumption that the vibration amplitude be small enough that the
particle is not inertially trapped at an antinode, and falls in the
typical parameter regime encountered in most experimental
situations. As an example, for a single Kr-atom on a $100$~MHz
CNT-resonator vibrating with an amplitude of $q_0=3$~nm at $T=1$~K,
one has $\omega_0{\cal E}/\gamma{\cal D}=E_{\rm vib}/k_{\mathrm
  B}T\approx 0.03$.

The corresponding Fokker-Planck equation (FPE) for the distribution function $p(\xi,E,\nu,\tau)$ reads~\cite{FPE}
\begin{eqnarray}
&&[\partial_\nu+\partial_\tau]{p}(\xi,E,\nu,\tau)=\epsilon\phi_0^2{\cal E}\sin(2\nu)\partial_{\cal E}p\nonumber \\
&&+\frac{\epsilon\phi_0^2}{2}[1+\cos(2\nu)]\partial_\nu p\nonumber\\ 
&&-\frac{\omega_0{\cal E}}{4\gamma}[1+\cos(2\nu)]\partial_\xi[p \partial_\xi \phi^2]+\frac{\cal D}{2}\partial_\xi^2 p.
\label{eq:FPE}
\end{eqnarray} 
As noted above, the dissipation of mechanical motion stems from the
correlation betwen the motion of the particle and the resonating beam,
induced by the last term in Eq.~(\ref{eq:FPE}). These correlations
occur on a time-scale $\omega_0^{-1}$ which is much shorter than the
scale of the rate of change of energy.  Hence, we can find the
dissipation rate by making a separation Ansatz for fast and slow time
scales by the approximation $p\approx p_0({\cal E},\tau)p_1({\cal
  E},\nu,\xi)$ (a more formal derivation is found in
Appendix~\ref{sec:perttheory}). This decouples Equation
~\eqref{eq:FPE} into one equation for slow time-scales and one for
fast time-scales, where particle position is described by the latter:
\begin{eqnarray}
\partial_\nu{p}_1=-\frac{\omega_0{\cal E}}{4\gamma}[1+\cos(2\nu)]\partial_\xi[p_1\partial_\xi \phi^2]+\frac{\cal D}{2}\partial_\xi^2 p_1. 
\end{eqnarray}

Assuming  $E_{\rm vib}/k_{\mathrm B}T \ll 1$ one finds to first order in $\cal{E}$ the
steady state solution
\begin{eqnarray}
&&{p_1}=1+\frac{\omega_0{\cal E}}{{\cal D}\gamma}\sum_{n}
  {\lambda_n f_{n}}\frac{\lambda_n\cos 2\nu+4\sin2
    \nu}{\lambda_n^2+16}\cos(n\pi \xi)\nonumber,
\label{eq:pe1}
\end{eqnarray}
where $\lambda_n={\cal D} n^2\pi^2$ and $f_{n}=\int_0^1
d\xi\,\cos(n\pi \xi)\phi_0^2(\xi)$. Inserting this solution into the
FPE equation~(\ref{eq:FPE}), and integrating over position $\xi$ and
the fast variable $\nu$ yields
\begin{eqnarray}
\partial_\tau{p_0}({\cal E},\tau)=\epsilon\frac{2\omega_0}{{\cal D}\gamma}\sum_{n} \frac{\lambda_n f_{n}^2}{\lambda_n^2+16}\partial_{\cal E}({\cal E}^2p_0).
\label{eq:alphaD}
\end{eqnarray}
The solution to this equation is 
$p({\cal E},\tau)={\cal E}^{-2}f\left(\frac{1-\alpha {\cal E}\tau}{\cal E}\right)$, where
\begin{eqnarray}
\alpha = \epsilon\frac{m\omega_0^2L^2}{k_{\mathrm B}T}\sum_{n}  f_{n}^2\frac{\lambda_n}{16+\lambda_n^{2}}.
\label{eq:alpha}
\end{eqnarray}

\begin{figure}[t]
\centering
\includegraphics[width=\linewidth]{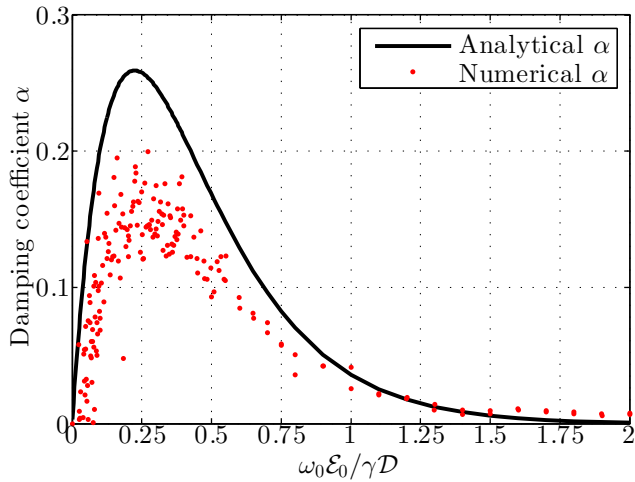}
\caption{(color online) Damping parameter $\alpha$, as calculated by
  Eq.~\eqref{eq:alpha} (solid line) and by a numerical fit
  (dots). The same simulation parameters as in Fig.~\ref{fig:decay}
  were used, with the exception of the initial amplitude; here $\mathcal
  E_0(0)=10^{-4}$.  The ratio $\omega_0\mathcal E_0/\gamma \mathcal D$
  was varied by changing the simulation temperature. We see that the
  perturbative approach is indeed valid for $\omega_0\mathcal
  E_0/\gamma\mathcal D\ll 1$. In the intermediate region $\omega_0
  \mathcal E_0/\gamma\mathcal D\lesssim 1$, Eq.~\eqref{eq:alpha}
  overestimates the magnitude of the damping, but captures the overall
  shape of the curve and the location of the maximal damping.}
\label{fig:alpha}
\end{figure}

If $p({\cal E} ,0)=\delta({\cal E}-{\cal E}_0)$ the ensemble averaged energy 
$\left<{\cal E}\right>$ decays without
dispersion and one obtains the characteristic ringdown of a
nonlinearly damped oscillator, 
\begin{eqnarray}
\left<{\cal E}(\tau)\right>=\frac{{\cal E}_0}{1+\alpha {\cal E}_0 \tau}.
\label{eq:ringdown}
\end{eqnarray}
This expression does indeed agree well with simulation in the parameter space where 
perturbation theory is valid; see Fig.~\ref{fig:alpha}.

To include dispersion, and to reach a proper thermal equilibrium
state, fluctuation corrections must be included. As shown in
Appendix~\ref{sec:perttheory}, this leads to the following Fokker-Planck equation
for the reduced probability density,
$$\partial_\tau p_0=\alpha\partial_{\cal E}{\cal E}^2[p_0+(\epsilon{\cal D}\gamma/\omega_0)\partial_{\cal E}p_0].$$
Noting that $(\epsilon{\cal D}\gamma/\omega_0)=2k_{\rm B}T/(M\omega_0^2L^2)$, we see 
that this FPE also gives the proper thermal equilibrium stationary solution $p_0(\tau\rightarrow\infty)\propto \exp(-M\omega_0^2q_0^2/2k_{\rm B}T)$. 

\section{Multimode dynamics, thermalization}
\label{sec:multimode}

\begin{figure}[t]
\centering
\includegraphics[width=\linewidth]{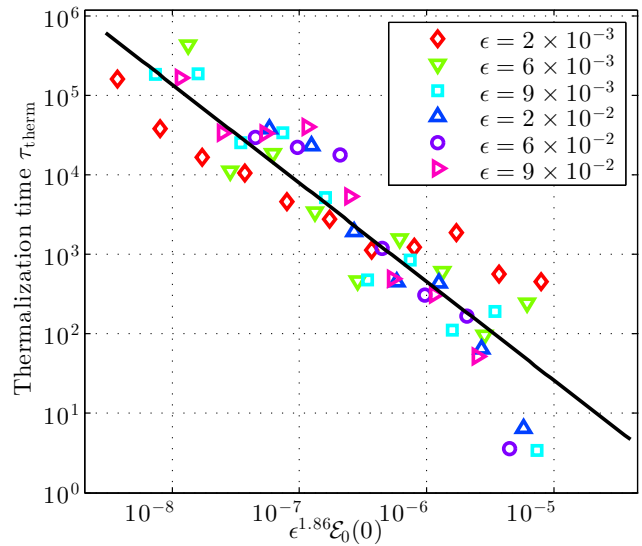}
\caption{(color online) Dependence of the thermalization time (the
  time at which $\mathcal E_n$ exceeds $k_{\mathrm B}T$ for some $n$)
  on $\epsilon$ and $\mathcal E_0(0)$. The black line is a
  least-squares fit to the data. The numerical simulation becomes more
  sensitive the more energy is put into the system, which explains the
  increased spread of the data points as $\mathcal E_0(0)$ increases.
\label{fig:tthermal}}
  
\end{figure}
While the qualitative behavior of the dynamics remain unchanged by
incorporating more flexural modes, a quantitative change takes place.
If the higher-lying modes are initially at rest, we find that exciting
the system in only the fundamental mode rapidly causes the higher
modes to be thermalized. Once in thermal equilibrium, they provide additional
channels for energy dissipation from the fundamental mode.

The thermalization of the higher modes stem from the stochastic
additive noise term in Eq.~(\ref{eq:eom1}). As initially ${\cal E}_0
\gg {\cal E}_{n>0}$, the transient behavior is described by
$$\ddot{q}_n+\omega_n^2 q_n\approx \epsilon \omega_0^2 q_0
\phi_n\phi_0\propto \epsilon\sqrt{{\cal E}_0},\quad n>0.$$ To a first
approximation, we would thus expect the energy of the higher-lying
modes to have a transient behavior $\left<{\cal E}_n\right> \sim
\epsilon^2{\cal E}_0t.$ In order to investigate this, we define 
the thermalization time $\tau_{\mathrm{therm}}$ as the time when the 
energy of an excited mode first exceeds the thermal energy. Simulations 
were made at a constant temperature but for values of $\epsilon$ and 
$\mathcal E_0(0)$ ranging over several orders of magnitude; the resulting 
values for $\tau_{\mathrm{therm}}$ are shown in Fig.~\ref{fig:tthermal}. 
The data has been fitted to a model $\tau_{\mathrm{therm}}\propto(\epsilon^a
\mathcal E_0)^b$, and the exponent $a=1.86$ ($b \approx 1.2$) was determined by minimizing 
the sum of squared residuals of the linear fit shown as a black line in 
Fig.~\ref{fig:tthermal}. The slight deviation from the theoretical value of 
$a=2$ is likely due to the present definition of $\tau_{\mathrm{therm}}$, which 
will always be smaller than the time taken for all higher-lying modes to 
reach $k_{\mathrm B}T$. 

The parameters $\epsilon$ and $\mathcal E_0(0)$ are straightforward to vary in experiments, e.g. by using different-sized nanoclusters as adsorbates and by varying the driving force used before beginning the ring-down. Hence, the linear dependence found in Figure~\ref{fig:tthermal} should be possible to verify experimentally.  

In addition, we note that for $\mathcal E_0(0)\gg 
k_{\mathrm B}T$, the thermalization of excited modes occur on a time scale 
much shorter than the decay of the fundamental mode. Consequently, on the time-
scale relevant for studying the ring-down of the resonator, it is a good 
approximation to assume that all excited modes are in thermal equilibrium.

The fact that the mode coupling strength and energy transfer between
modes are determined only by the energy in the modes is further
corroborated by considering the decay rate of the fundamental mode
energy once the higher
modes have thermalized. As shown in Fig. 6, for each additional mode
we include in the simulation, an additional channel for energy
transfer away from the fundamental mode is made available and the
decay rate (initially) increases linearly with the number of added
modes. As each individual mode has the same energy $\sim k_{\rm B}T$,
each mode contributes an equal amount to the fundamental
mode dissipation. 

\begin{figure}[t]
\centering
\includegraphics[width=\linewidth]{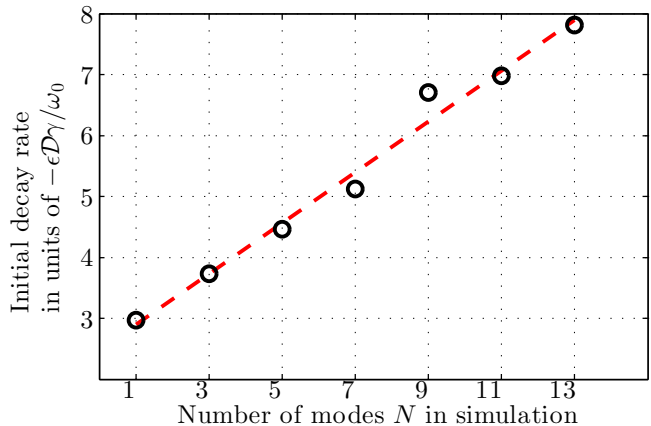}
\caption{(color online) Slope of the decay of the fundamental mode during the initial, 
  linear regime, as a function of the number of modes $N$ included in the simulation; 
  black circles are data points while the red dashed line is a linear fit.  The values 
  have been divided by $-\epsilon{\cal D}\gamma/\omega_0$ to show the agreement with 
  the theoretical case~\eqref{eq:linearE}. As more excited modes are included, more 
  decay channels are opened and the rate of decay increases. 
\label{fig:Ndep}}
\end{figure}

Clearly, there must be a cut-off at which this is no longer true. Such
a cutoff can be estimated by noting that due to the influence of the
fundamental mode, the dissipation of energy from the fundamental mode
is associated with adsorbate dynamics occuring on a timescale
$\omega_0^{-1}$ corresponding to a diffusion length scale of
$\sqrt{2\pi D/\omega_0}$. If this length is larger than half the
wavelength of the $n^{\mathrm{th}}$ mode, the effective coupling to
this mode will average to zero and not contribute to
dissipation. The wavelength is $\lambda_n=2L/(n+1)$, giving the
cut-off condition that only modes with $n\lesssim n_{\rm max}\approx
\sqrt{\frac2{\pi\mathcal D}}$ will contribute to the dissipation of
the fundamental mode. For the parameter values used in
Fig.~\ref{fig:Ndep} we find $n_{\rm max}\approx 20$.

%%%%%%%%%%%%%

\section{Conclusions}	
We have studied the effect of a diffusing adsorbed particle on a
vibrating 1D nanomechanical resonator, initially excited in its
fundamental flexural mode. Studying the free ring-down of the mode,
focusing on the energy transfer induced by the diffusion 
we find that there are two effects that cause vibrational energy to dissipate.

First, the inertial force exerted on the adsorbed particle causes
short-time correlations between adsorbate motion and flexural
vibrations. For large initial amplitudes such that
$m\omega_0^2q_0^2/k_{\rm B}T\gg 1$, the particle is trapped at the
antinode of vibration, and the decay of vibrational energy is linear in
time, approximately given by expression~\eqref{eq:linearE}. For lower
amplitudes, when the particle diffuses freely along the resonator, the
decay takes the same form as that of a nonlinearly damped oscillator
according to Equation~\eqref{eq:ringdown}.

Second, the diffusing particle also provides a stochastic coupling
between different flexural vibration modes. This stochastic coupling
provides an additional channel for energy transfer from the
fundamental mode for each added flexural mode. This second mechanism
can further significantly lower the ring-down time of the fundamental
mode and causes rapid thermalization among the higher modes. The dissipation rate due to the mode coupling is linear in the number of included excited modes (up to a parameter-dependent cut-off), meaning that a single mode-treatment of a resonator is qualitatively but not quantitatively correct. The exact thermalization rate of the excited modes depends on adsorbate mass ($\epsilon$), device geometry ($\omega_0$), and initial amplitude ($\mathcal E_0(0)$); parameters that are readily accessible in experiments. 

With recent advances in readout of the real-time evolution of
nanomechanical oscillators~\cite{Meerwaldt_2013, Leeuwen_2014}
together with the ability to deposit individual particles on
ultrahigh-Q carbon nanotube resonators~\cite{Bachtold_2012}, the
proposed effects should be possible to observe experimentally. 
These
result also have bearing on the numerical analysis in
Ref.~\onlinecite{Park_2014} which showed a dramatic change in Q-factor
for an Au-cluster deposited on a graphene resonator at the onset of
particle diffusion. Finally, the existence of a trapping regime where adsorbate diffusion is suppressed implies that mass sensing experiments above cryogenic temperatures may be possible, given that the resonator is driven strongly enough.

%%%%%%%%%%%%%%%%%%%%%%%%%%%%%%%%%%%%%%%%%%%%%%%%%%%%%%%%%%%%%%%%%%%%%		
\begin{acknowledgments}
We acknowledge financial support from the Swedish Research Council VR
(AI), the The Foundation for Strategic  Research SSF (CE) as well as the
European Union through grant no. 246026 (AI, CE) and the GRAPHENE
Flagship (CE).
\end{acknowledgments}

%%%%%%%%%%%%%%%%%%%%%%%%%%%%%%%%%%%%%%%%%%%%%%%%%%%%%%%%%%%%%%%%%%%%%%%%%%%%%
\appendix

\section{Determining the $\boldsymbol{\omega_n}$ and $\boldsymbol{\phi_n(X)}$ 
\label{sec:eigenmodes}}
Here, we derive the flexural Eigenmodes and Eigenfrequencies for the unperturbed resonator.

The equation of motion corresponding to the Lagrangian~\eqref{eq:unpertL} is
\begin{eqnarray}
&&\rho\ddot w-\sigma \partial_X^2w+\kappa \partial_X^4w=0.
\end{eqnarray}
Defining $\phi_n(X)$ and $\omega_n$ through
$w(X,t)=e^{-i\omega_nt}\phi_n(X)$, we find that the Eigenmodes satisfy
the equation
\begin{eqnarray}
&&-\rho\omega_n^2\phi_n-\sigma \phi_n''+\kappa \phi_n''''=0.
\end{eqnarray}
The corresponding characteristic equation is
\begin{eqnarray}
&&\kappa k^4-\sigma k^2-\rho\omega_n^2=0
\end{eqnarray}
with roots $\pm k_n^+,\ \pm ik_n^-$, where
\begin{eqnarray}
&&k_n^\pm=\sqrt{\sqrt{\frac{\sigma^2}{4\kappa^2}+\frac{\rho\omega_n^2}{\kappa}}\pm\frac{\sigma}{2\kappa}},\ n=0,1,2,\ldots
\end{eqnarray}
Hence, the Eigenmodes can be written as
\begin{eqnarray}
\phi_{2n} &=& A_{2n}\ch k_{2n}^+(X-\tfrac L2)+A_{2n}'\cos k_{2n}^-(X-\tfrac L2)\nonumber\\
\phi_{2n+1} &=& B_{2n+1}\sh k_{2n+1}^+(X-\tfrac L2)\nonumber\\
&+&B_{2n+1}'\sin k_{2n+1}^-(X-\tfrac L2),
\end{eqnarray}
where the boundary conditions $\phi_n(0)=\phi_n(L)=\phi_n'(0)=\phi_n'(L)=0$ have 
been used to divide the $\phi_n$ into even and odd sets of Eigenmodes. Similarily, 
we find that the Eigenfrequencies $\omega_n$ are determined from the equation
\begin{eqnarray}
\frac{k_{n}^\mp}{k_{n}^\pm}=\pm\frac{\tan k_{n}^-\tfrac L2}{\tanh k_{n}^+\tfrac L2},
\end{eqnarray}
where upper/lower signs correspond to odd/even $n$.
A good approximation for the ratio $\omega_n/\omega_0$ is $(2n+1)^2/9$.  
Finally, the integration constants $A_n$ and $B_n$ are determined. The boundary 
conditions demand that
\begin{eqnarray}
&&A_n'=-\frac{\cosh k_n^+\tfrac L2}{\cos k_n^-\tfrac
    L2}A_n,\ \ B_n'=-\frac{\sinh k_n^+\tfrac L2}{\sin k_n^-\tfrac
    L2}B_n
\end{eqnarray}
whereas the normalization condition $\int \mathrm d X \phi_m^\dagger\phi_n = 
L\delta_{mn}$ determines
\begin{eqnarray}
&&|A_n|^2=2\Bigg[ 1+\frac{\sh
      k_{n}^+L}{k_{n}^+L}+\frac{\ch^2k_{n}^+\tfrac
      L2}{\cos^2k_{n}^-\tfrac L2}\left(1+\frac{\sin
      k_{n}^-L}{k_{n}^-L}\right)\Bigg]^{-1},\nonumber\\ 
&&|B_n|^2=2\Bigg[
    1-\frac{\sh k_{n}^+L}{k_{n}^+L}-\frac{\sh^2k_{n}^+\tfrac
      L2}{\sin^2k_{n}^-\tfrac L2}\left(1-\frac{\sin
      k_{n}^-L}{k_{n}^-L}\right)\Bigg]^{-1}.\nonumber\\
\end{eqnarray}
The final undetermined phase is chosen so that the Eigenfunctions $\phi_n$ are real.

%%%%%%%

\section{Formal perturbation theory \label{sec:perttheory}}
The perturbation theory sketched in section~\ref{sec:whatever} can be
put on more formal grounds. In this appendix we derive the reduced FPE
by means of the methods in Ref.~\onlinecite{FPE}.  Introducing the variable
$\nu=\omega_0 t+\theta$, the FPE reads
\begin{eqnarray}
&&\partial_\tau p=\epsilon\phi_0^2{\cal E}\sin(2\nu)\partial_{\cal
    E}p\nonumber
  \\ &&\left(\frac{\epsilon\phi_0^2}{2}[1+\cos(2\nu)]-1\right)\partial_\nu
  p\nonumber\\ &&-\frac{\omega_0{\cal
      E}}{2\gamma}[1+\cos(2\nu)]\partial_\xi[p \partial_\xi
    \phi^2]+{\cal D}\partial_\xi^2 p.
\label{eq:FPE2}
\end{eqnarray}
Upon expanding $p=\sum_n p_n({\cal{E}},\xi,\tau)e^{2in\nu}$ and
introducing the vector ${\bf p}=[..,p_1,p_0,p_1,..]^\bot$, the
FPE-equation can be rewritten as
$$\partial_\tau{\bf p}=\epsilon \hat L_1{\bf p}+\frac{\cal D}{2}\hat
L_2{\bf p},$$ where $\hat{L}_1{\bf p}=\phi_0^2[{\cal E}\hat A
\partial_{\cal E}+i\hat{B}\hat{N}]{\bf p}$ and
$$\hat{L}_2{\bf p}=-4i{\cal D}^{-1}\hat N{\bf p}-\frac{\omega_0{\cal
    E}}{2{\cal D}\gamma}\hat{B}\partial_\xi[{\bf p} \partial_\xi
  \phi^2]+\partial_\xi^2 {\bf p}.$$ The matrices $\hat{A}$, $\hat{B}$,
$\hat{N}$, have components
$A_{m,n}=(2i)^{-1}[\delta_{m,n+1}-\delta_{m,n-1}]$,
$B_{m,n}=\delta_{m,n}+2^{-1}[\delta_{m,n+1}+\delta_{m,n-1}]$, and
$N_{m,n}=n\delta_{m,n}$.

Expanding in Eigenmodes of the operator $\hat{L}_2$, as ${\bf
  p}=\sum_n \beta_n({\cal E,\tau}){\bf v}_n(\xi,{\cal E})$, where
$\hat{L}_2{\bf v}_n=-\mu_n({\cal E}){\bf v}_n$ yields
$$\left(\partial_\tau+\frac{{\cal
    D}\mu_n}{2}\right)\beta_n=\epsilon\sum_m\left< {\bf w}_n,
\hat{L}_1 \beta_m {\bf v}_m\right>.$$ The inner product is here
defined as $\left<{\bf u},{\bf v}\right>\equiv \int d\xi\, {\bf
  u}^\dagger {\bf v}$ and the right eigenvectors ${\bf w}_n$ satisfy
the adjoint equation $L_2^{*}{\bf w}_n=-\mu_n^*{\bf w}_n$ with
$$L_2^{*}=4i{\cal
  D}^{-1}\hat N+\frac{\omega_0{\cal E}}{2{\cal D}\gamma}\hat{B}(\partial_\xi \phi_0^2)\partial_\xi+\partial_\xi^2.$$
For time scales $\tau>{\cal D}^{-1}$, we can make the approximation
\begin{eqnarray}
&&\partial_\tau\beta_0=\epsilon\sum_n\left< {\bf w}_0, \hat{L}_1 \beta_n {\bf v}_n\right>,\nonumber\\
&&\beta_{n\ge 1}\approx\frac{2\epsilon}{{\cal D}\mu_n}\left< {\bf w}_n, \hat{L}_1 \beta_0 {\bf v}_0\right>.
\end{eqnarray}
Combining the two gives
\begin{eqnarray}
&&\partial_\tau\beta_0=\epsilon\left< {\bf w}_0, \hat{L}_1 \beta_0 {\bf v}_0\right>\nonumber\\
&&+\frac{2\epsilon^2}{{\cal D}}\sum_{n\ge 1}\left< {\bf w}_0,  \hat{L}_1 \left(\mu_n^{-1}\left< {\bf w}_n, \hat{L}_1 \beta_0 {\bf v}_0\right>\right) {\bf v}_n\right>\nonumber\\
\end{eqnarray}

%%%%%%%

\subsection{Perturbation theory for Eigenvectors}
The Eigenvectors of ${\hat L}_2$ cannot obtained exactly. However, if
the parameter $\eta\equiv \frac{\omega_0{\cal E}}{2{\cal D}\gamma}\ll
1$ we can find them perturbatively to first order in $\eta$.  Each
Eigenvector -- Eigenvalue pair has composite indices $(n,m)$ and is to
first order given by
\begin{eqnarray}
&&{\bf v}_{nm}\approx {\bf v}_{nm}^{(0)}+\frac{\omega_0{\cal
      E}}{2{\cal D}\gamma}\sum_{pq\neq mn}\frac{\left<{\bf
        w}_{pq}^{(0)},\hat{B}\partial_\xi{\bf
      v}_{nm}^{(0)}\partial_\xi\phi_0^2\right>}{\mu_{nm}^{(0)}-\mu_{pq}^{(0)}}{\bf
    v}_{pq}^{(0)}\nonumber\\ 
&&{\bf w}_{nm}\approx {\bf w}_{nm}^{(0)}-\frac{\omega_0{\cal
      E}}{2{\cal D}\gamma}\sum_{pq\neq mn}\frac{\left<{\bf
        w}_{pq}^{(0)},\hat{B}(\partial_\xi\phi_0^2)\partial_\xi{\bf
      v}_{nm}^{(0)}\right>}{(\mu_{nm}^{(0)}-\mu_{pq}^{(0)})^*}{\bf
    w}_{pq}^{(0)}\nonumber\\ 
&&\mu_{nm}\approx
  \mu_{nm}^{(0)}+\frac{\omega_0{\cal E}}{2{\cal
      D}\gamma}\left<[{\bf
      w}_{nm}^{(0)}],\hat{B}\partial_\xi{\bf
    v}_{nm}^{(0)}\partial_\xi\phi_0^2\right>.\nonumber
\end{eqnarray}
The unperturbed eigenvalues are $\mu_{nm}^{(0)}=n^2\pi^2+4i{\cal
  D}^{-1}m$ and the corresponding eigenvector has the $k$:th component
$$[{\bf v}_{nm}^{(0)}]_k=[{\bf w}_{nm}^{(0)}]_k=[\sqrt{2}\cos(n\pi\xi)+\delta_{n,0}(1-\sqrt{2})]\delta_{k,m}.$$ 

%%%%%%%

\subsection{Derivation of FPE for reduced density $\boldsymbol{\beta_0({\cal E},\tau)}$}
To obtain the FPE for the slowly varying coefficient $\beta_0({\cal E},\tau)$, corresponding to the probability density $p_0({\cal E},\tau)$ in the main text, we first observe that  
\begin{eqnarray}
&&\left<{\bf w}_0,{\cal L}_1 F({\cal E}){\bf v}_{nm}\right>=\nonumber\\
&&(i/2)\partial_{\cal E}\left[{\cal E}F({\cal E})\int d\xi \phi_0^2([{\bf v}_{nm}]_1-[{\bf v}_{nm}]_{-1})\right]
\end{eqnarray}
for an arbitrary function $F({\cal E})$. With $F({\cal
  E})=\beta_0({\cal E})$ we then recover the expression in
Eq.(\ref{eq:pe1}), 
$$\epsilon\left<{\bf w}_0, \hat{L}_1 \beta_0 {\bf v}_0\right>=\epsilon\frac{2\omega_0}{{\cal D}\gamma}\sum_n f_n^2\frac{\lambda_n}{(\lambda_n)^2+16}\partial_{\cal E}[{\cal E}^2\beta_0].$$

\begin{figure}[t]
\includegraphics[width=\linewidth]{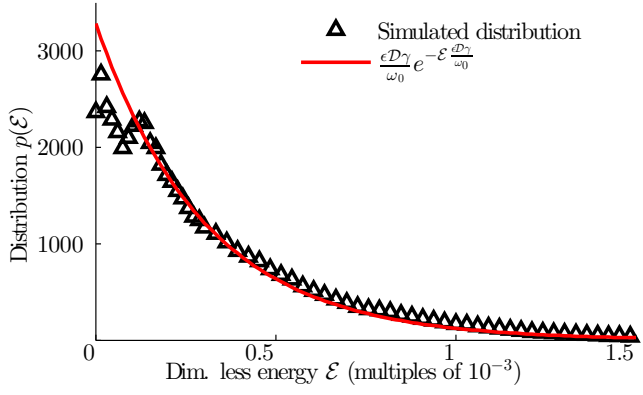}
\caption{(color online) Equilibrium distribution $p_{\rm st.}({\cal
    E})$. (Black triangles) Distribution obtained from numerical
  simulation of Equations.~\eqref{eq:actang1}-\eqref{eq:actang3} after
  relaxation. (Red line) The distribution $p({\cal E})=(\epsilon {\cal
    D}\gamma/\omega_0)\exp(-{\cal E}(\epsilon {\cal
    D}\gamma/\omega_0)).$\label{fig:eqdistr} }
\end{figure}

For the fluctuation correction, the lowest order term arises from
considering only the unperturbed eigenvectors, 
\begin{eqnarray}
&&2\epsilon^2\left< {\bf w}_0^{0},  \hat{L}_1 \left((\mu_{nm}^{0})^{-1}\left< {\bf w}_{nm}^{0}, \hat{L}_1 \beta_0 {\bf v}_0^{0}\right>\right) {\bf v}_{nm}^{0}\right>\nonumber\\
&&=2\epsilon^2\sum_n\frac{f_n^2\lambda_n}{\lambda_n^2+16}\partial_{\cal E}{\cal E}^2\partial_{\cal E}\beta_0=\alpha\left(\frac{\epsilon{\cal D}\gamma}{\omega_0}\right)\partial_{\cal E}{\cal E}^2\partial_{\cal E}\beta_0.\nonumber
\end{eqnarray}
Hence, to lowest order in $\cal E$ and to second order in $\epsilon$, the FPE reads
$$\partial_\tau \beta_0=\alpha\partial_{\cal E}{\cal E}^2\left(\beta_0+\left(\frac{\epsilon{\cal D}\gamma}{\omega_0}\right)\partial_{\cal E}\beta_0\right).$$
The stationary solution to this equation is the equilibrium distribution $p({\cal E})=(\epsilon {\cal D}\gamma/\omega_0)\exp(-{\cal E}(\epsilon {\cal D}\gamma/\omega_0))$ which we also confirm by direct numerical simuation (see Fig.~\ref{fig:eqdistr}).

%%%%%%%%%%%%%%%%%%%%%%%%%%%%%%%%%%%%%%%%%%%%%%%%%%%%%%%%%%%%%%%%%%%%%
%%%%%%%%%%%%%%%%%%%%%%%%%%%%%%%%%%%%%%%%%%%%%%%%%%%%%%%%%%%%%%%%%%%%%		
		
%%	BIBLIOGRAPHY

%\bibliography{nldmgrex}

\begin{thebibliography}{99}
\bibitem{Roukes_2011} J. L. Arlett, E. B. Myers, and M. L. Roukes, Nature Nanotechnology {\bf 6}, 203 (2011).
\bibitem{Roukes_2012} M. S. Hanay, S. Kelber, A. K. Naik, D. Chi, S. Hentz, E. C. Bullard, E. Colinet, L. Duraffourg, and M. L. Roukes, Nature Nanotechnology {\bf 7}, 602 (2012). 
\bibitem{Bachtold_2008} B. Lassagne, D. Garcia-Sanchez, A. Aguasca, and A. Bachtold, Nano Letters {\bf 8}, 3735 (2008).
\bibitem{Zettl_2008} K. Jensen, K. Kim, and A. Zettl, Nature Nanotechnology, {\bf 3}, 533 (2008). 
\bibitem{Bachtold_2012} J. Chaste, A. Eichler, J. Moser, G. Ceballos, R. Rurali, and A. Bachtold, Nature Nanotechnology {\bf 7}, 301 (2012).
\bibitem{Atalaya_2011} J. Atalaya, A. Isacsson, and M. I. Dykman, Physical Review B {\bf 83}, 045419 (2011).
\bibitem{Dykman_2010} M. I. Dykman, M. Khasin, J. Portman, and S. W. Shaw, Physical Review Letters {\bf 105}, 230601 (2010).
\bibitem{Atalaya_2011_2} J. Atalaya, A. Isacsson, and M. I. Dykman, Physical Review Letters {\bf 106}, 227202 (2011). 
\bibitem{Atalaya_2012} J. Atalaya, J. Phys. C Cond. Mat. {\bf 24}, 475301 (2012).
\bibitem{Roukes_Diffusion} Y. T. Yang, C. Callegari, X. L. Feng, and M. L. Roukes, Nano Letters {\bf 11}, 1753 (2011).
\bibitem{Lifshitz_2000} R. Lifshitz and M. L. Roukes, Phys. Rev. B {\bf 61}, 5600 (2000).
\bibitem{Lifshitz_2001} M. C. Cross and R. Lifshitz, Phys. Rev. B {\bf 64}, 085324 (2001).
\bibitem{WilsonRae_2008} I. Wilson-Rae, Phys. Rev. B {\bf 77}, 245418 (2008).
\bibitem{Eom_2011} K. Eom, H. S. Park, D. S. Yoon, and T. Kwon, Phys. Rep. {\bf 503}, 115 (2011).
\bibitem{Eichler_2011} A. Eichler, J. Moser, J. Chaste, M. Zdrojek, I. Wilson-Rae, and A. Bachtold, Nat. {\bf 6}, 339 (2011).
\bibitem{Croy_2012} A. Croy, D. Midtvedt, A. Isacsson, and J. M. Kinaret, Phys. Rev. B {\bf 86}, 235435 (2012). 
\bibitem{Midtvedt_2014} D. Midtvedt, A. Croy, A. Isacsson, Z. Qi, and H. S. Park, Phys. Rev. Lett. {\bf 112}, 145503  (2014).
\bibitem{McEuen_PNAS} A. W. Barnard, V. Sazonova, A. M. van der Zande, and P. L. McEuen, PNAS {\bf 109}, 19093 (2012). 
\bibitem{Park_2014} J.-W. Jiang, B.-S. Wang, H. S. Park,  et al., Nanotechnology {\bf 2}, 02501 (2014).
\bibitem{Meerwaldt_2013} H. B. Meerwaldt, S. R. Johnston, H. S. J. van der Zant, and G. A. Steele, Appl. Phys. Lett. {\bf 103}, 053121 (2013). 
\bibitem{Leeuwen_2014} R. van Leeuwen, A. Castellanos-Gomez, G. A. Steele, H. S. J. van der Zant, W. J. Venstra, Appl. Phys. Lett. {\bf 105}, 041911 (2014).
\bibitem{LL} L. D. Landau and E. M. Lifshitz,
  \emph{"Theory of Elasticity"}, Third Edition,
        ELSEVIER BUTTERWORTH HEINMANN, New York, (1986).
\bibitem{Manella_1989} R. Mannella and V. Palleschi, Phys. Rev. A {\bf 40}, 3381 (1989).
\bibitem{FPE} H. Risken, "The Fokker-Planck Equation: Methods of Solution and Applications (Springer Series in Synergetics)", 2nd Ed. Springer Verlag, New York. (1996)

\end{thebibliography}

\end{document}